\documentclass[preprint,aps,showpacs,floatfix]{revtex4}
\usepackage{amsmath,amssymb,graphicx,epsfig}

\newcommand{\vn}{v_{\rm n}}
\newcommand{\Ree}{{\rm {Re}}_D}
\newcommand{\mun}{\mu_{\rm n}}
\newcommand{\xxo}{x_1^0}
\newcommand{\yyo}{y_1^0}

\newcommand{\zz}{z_1}
\newcommand{\zzc}{z_1^*}
\newcommand{\zzz}{z_2}
\newcommand{\zzzc}{z_2^*}
\newcommand{\zzo}{z_1^0}
\newcommand{\zzzo}{z_2^0}
\newcommand{\zzoc}{(z_1^0)^*}

\begin{document}
%\draft
%\def\pd#1#2{\frac{\partial #1}{\partial #2}}

\title{\bf Normal fluid eddies in the thermal counterflow past
a cylinder}

\author{Y. A. Sergeev${}^1$ and C. F. Barenghi${}^2$}

\affiliation { ${}^1$School of Mechanical and Systems Engineering,
Newcastle University, Newcastle upon Tyne NE1 7RU, United
Kingdom\\ ${}^2$School of Mathematics and Statistics, Newcastle
University, Newcastle upon Tyne NE1 7RU, United Kingdom}
\date {\today}

\begin {abstract}
A recent Particle Image Velocimetry (PIV) experiment in He~II
counterflow around a cylindrical obstacle showed the existence of
apparently stationary normal fluid eddies both downstream (at the
rear) and upstream (in front) of the cylinder. This rather
surprising result does not have an analogue in experimental
observations of classical fluid flows. We suggest that the
explanation for the apparent stability of such eddies can be
provided entirely from the viewpoint of classical fluid dynamics.
We also discuss a possible connection between the emergence of the
normal fluid eddies and the polarization of the vortex tangle in
superfluid.
\end{abstract}

\pacs{
67.40.Vs, 47.37.+q, 47.27.-i \\}

\maketitle

\section{Motivation} \label{Motivation}
Particle Image Velocimetry (PIV), which has been, for several
decades, a standard technique of flow visualization in classical
fluid dynamics, is based on tracking the motion of solid particles
injected into the fluid. Provided the particles are sufficiently
small, they can be expected to follow faithfully the fluid flow.

Recently implemented in helium~II \cite{PIV}, the PIV technique
has already yielded some non-trivial and unexpected
results~\cite{VanSciver,VanSciverNat,Bewley,Paoletti} followed by
the attempts of their theoretical interpretation (see e.g.
Refs.~\cite{Poole,Sergeev}). Among surprising experimental results
is the recent observation~\cite{VanSciverNat} of the apparently
stationary normal fluid eddies in the thermal counterflow past a
cylinder. In the cited work, Zhang and Van Sciver visualized the
motion of small particles in the thermal counterflow around the
cylinder of diameter $D=0.635\,{\rm cm}$ fixed in the center of
rectangular channel of a cross-section $3.89\times1.95\,{\rm
cm}^2$. The counterflow was produced, in two separate experiments,
by the heat flux $q=0.4$ and $1.12\,{\rm W/cm^2}$ at temperatures
$T=1.6$ and $2.03\,{\rm K}$, respectively (corresponding to the
Reynolds numbers $\Ree=\rho D\vn/\mun=4.1\times10^4$ and
$2.1\times10^4$, where $\rho=0.145\,{\rm g/cm^3}$ is the density
of the liquid helium, and $\vn$ and $\mun$ are the mean velocity
and the viscosity of the normal fluid). Solid particles used for
visualization in the PIV experiments were polymer microspheres of
diameter $1.7\,\mu{\rm m}$ and density $1.1\,{\rm g/cm^3}$.

In these experiments Zhang and Van Sciver observed the formation of
large-scale eddies of the particulate motion located both downstream
and, surprisingly, upstream of the cylinder with respect to the
normal flow (it should be noted that these structures were somewhat
more discernible in the second of two experiments, i.e for $\Ree
=2.1\times10^4$, in which case the corresponding mean normal fluid
velocity was $\vn\approx2.2\,{\rm cm/s}$). These, apparently stable
vortices of the particulate flow field were located at distances
about 3 cylinder radii from its center at the angles $\pm45^o$ and
$\pm135^o$ to the axis along the undisturbed flow through the center
of the cylinder.

In order to interpret these observations, the following question
should first be addressed: what do tracer particles actually
trace? It seems natural to assume that in the cited experiment, because
the Stokes drag of small particles is much larger than other forces
exerted by the normal fluid (see e.g. Ref.~\cite{Poole}), the
particle trace the normal fluid. However, solid particles also
interact strongly with quantized vortices which may reconnect to the
particle surface; such interactions may lead to the appearance of
the additional force exerted on particles by the superfluid. This
additional force was actually measured in another Zhang and Van
Sciver's experiment~\cite{VanSciver} on particle sedimentation in
the thermal counterflow, and later analyzed
theoretically~\cite{Sergeev} by the authors of the present work.

A second question is: do the circulation cells of the particulate
motion map, in the experiment~\cite{VanSciverNat}, the normal
eddies, or result from complex interactions of particles with both
the normal fluid and quantized vortices in the superfluid
component (in which case the eddies in the normal fluid may not
even exist)? Since in this experiments the vortex tangle was
relatively dilute (see more detaled discussion below in
Sec.~\ref{interpretation}), we can expect with some confidence
that the particle motion indeed maps, at least qualitatively, the
normal flow. The observed normal flow patterns do not have a
classical analogue: in classical fluid dynamics, stationary eddies
upstream of the cylinder have never been observed, and downstream
of the cylinder one would rather expect the Von K\'arm\'an vortex
street than a pair of apparently stationary eddies. In
Ref.~\cite{VanSciverNat} it was suggested that the observed large
normal vortex structures were caused by the complex interaction
between the two fluid components of He~II.

In the present work we argue that, while the emergence of the
large-scale vortex structures is, most likely, caused by the mutual
friction between quantized vortices and the normal fluid, the
apparent stability of the observed eddies, both at the rear
and in front
of the cylinder, can be explained entirely from the viewpoint of
classical fluid dynamics without invoking an interaction between the
normal fluid and superfluid vortices.

Because the Reynolds numbers in the cited experiment were large
($\sim10^4$), instead of considering the turbulent viscous flow
past a cylinder we will analyze a much simpler model of the plane
motion of two point vortices of opposite polarities in the
inviscid, potential flow around a disk. We shall show that there
exist stationary configurations of the vortex-antivortex pair,
both behind and in front of the disk. Such configurations are
unstable, i.e. any perturbation of the stationary configuration
leads to eventual sweeping of the vortex points away from the
disk. However, we shall show that there exist stationary
configurations such that, on the time scale corresponding to the
duration of experiment~\cite{VanSciverNat}, the vortices located
sufficiently close to the corresponding stationary points will
remain in the close proximity of their initial locations.
Moreover, such stationary points are positioned as the apparently
stable eddies seen by Zhang and Van Sciver. In conclusion we shall
also discuss a possible connection between the emergence of the
normal fluid eddies and the polarization of the vortex tangle in
the superfluid component of He~II.

\section{Lagrangian equations of motion of point vortices in the
inviscid flow around the disk} \label{motion}

\begin{figure}
\begin{tabular}[b]{cc}
\includegraphics[height=0.33\linewidth]{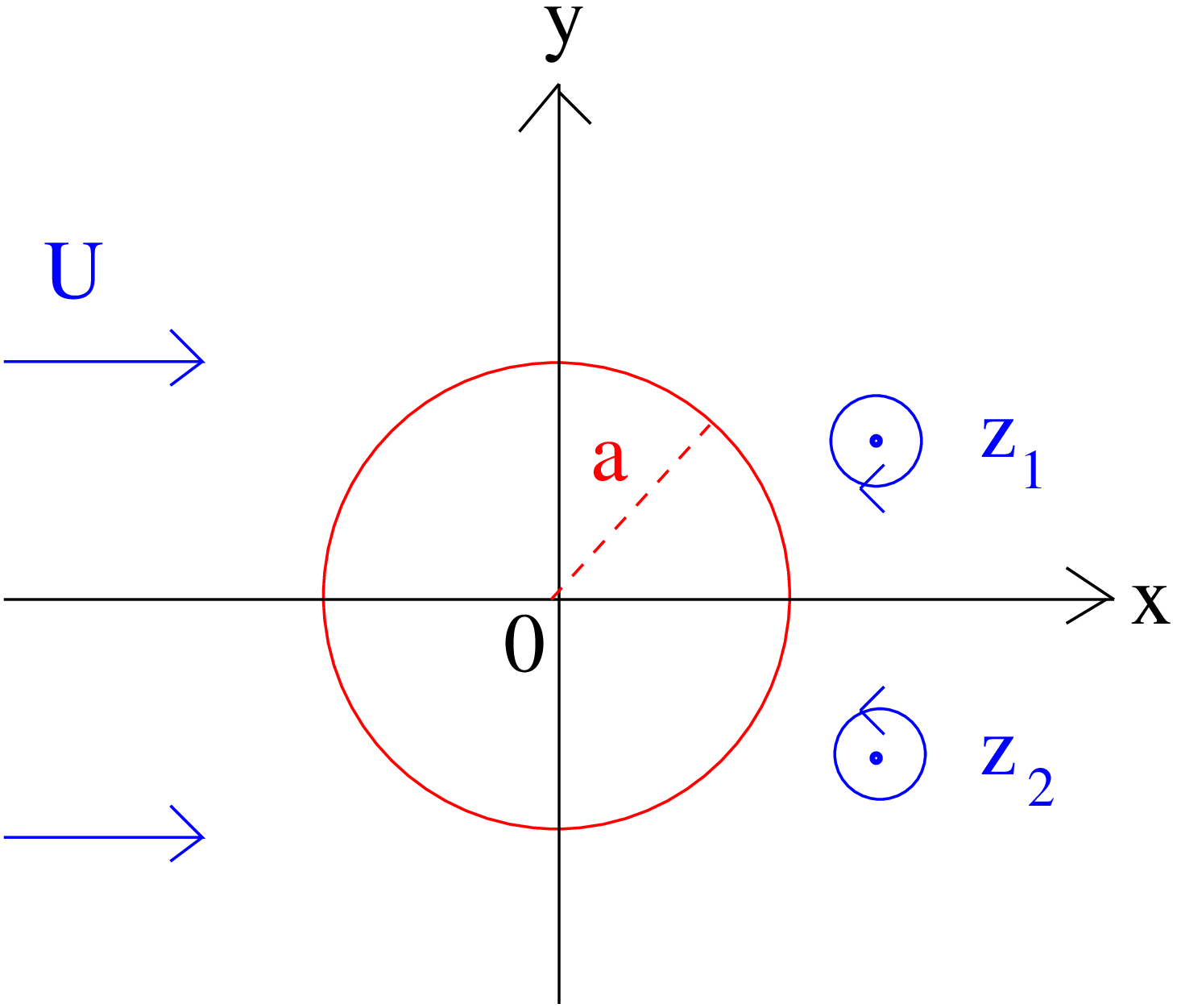}&
\includegraphics[height=0.33\linewidth]{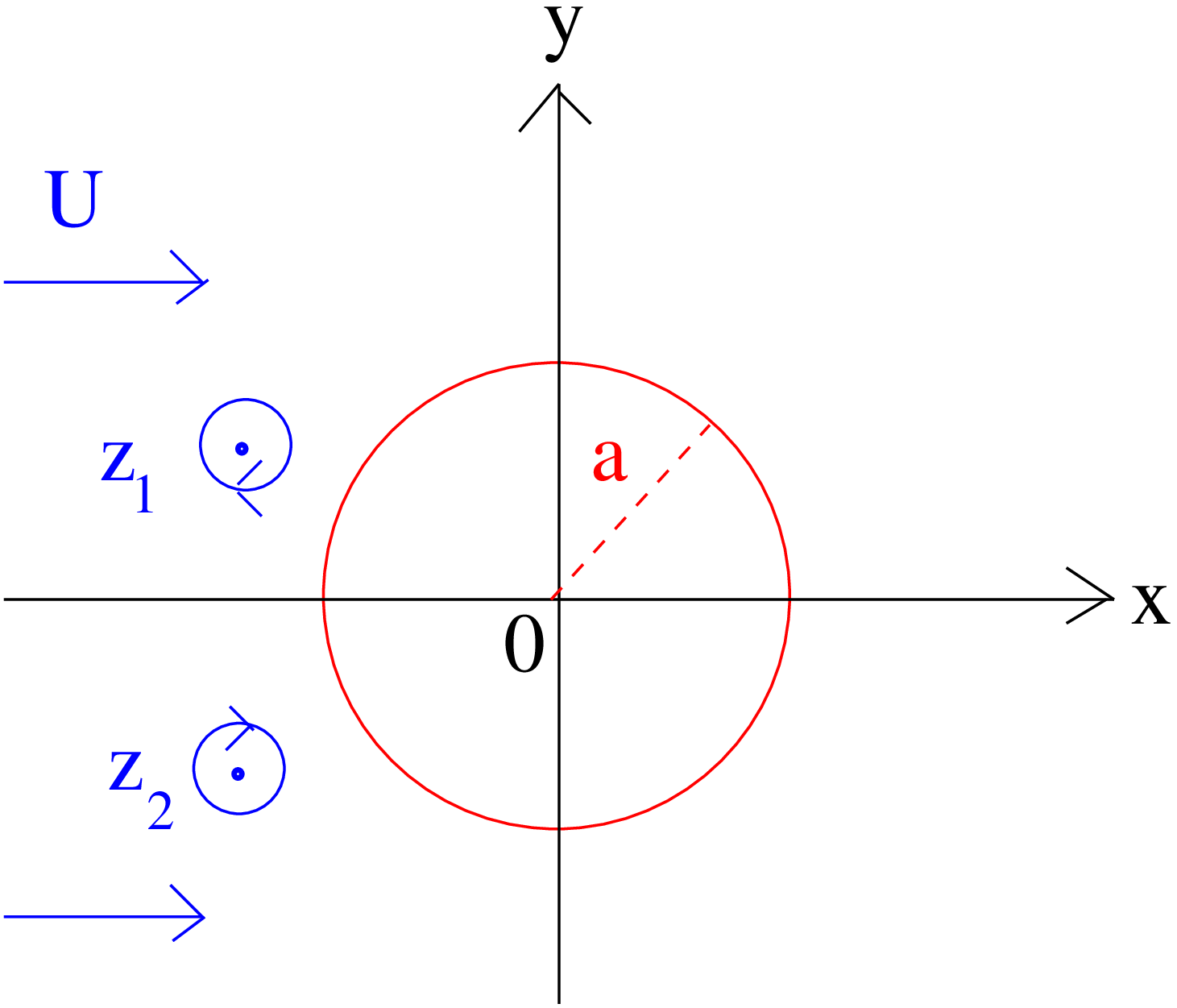}\\
\end{tabular}
\caption{(Color online). Point vortices in the potential flow
around the disk. Left: vortex-antivortex pair downstream (at the
rear) of the disk. Right: vortex-antivortex pair upstream (in
front) of the disk.} \label{fig:1}
\end{figure}

We consider the two-dimensional inviscid, potential flow, with
velocity $U$ at infinity, around a circular disk of radius $a$. We
are concerned with a motion in such a flow of a vortex-antivortex
pair, i.e. of two vortices of opposite polarities but the same
circulation, $\Gamma$. Let the positions of the vortex points on
the complex plane be $\zz(t)=x_1(t)+iy_1(t)$ and
$\zzz(t)=x_2(t)+iy_2(t)$, see Fig.~\ref{fig:1}. The reason why the
vortex $\zz$ has negative (clockwise) circulation, $\Gamma<0$,
will become clear in Sec.~\ref{Stationary}.

Each of these vortices moves as a fluid point in the superposition
of the imposed flow around the disk and the flow field of another
vortex. Complex potential of the superposition of the potential,
uniform at infinity, flow around the disk and the flow created by
the vortex point at, say, $z=z_1$ is \cite{Batchelor}
\begin{equation}
w(z)=U\biggl(z+\frac{a^2}{z}\biggr)+\frac{\Gamma}{2\pi
i}\biggl[\ln(z-\zz)-\ln\biggl(\frac{a^2-z\zzc}{z}\biggr)\biggr]\,,
\label{potential}
\end{equation}
where $z^*=x-iy$ denotes the complex conjugate of $z=x+iy$.

It is convenient to introduce the non dimensional variables
\begin{equation}
z'=\frac{z}{a}\,, \quad w'=\frac{w}{Ua}\,, \quad
\lambda=-\frac{\Gamma}{4\pi Ua}>0, \label{nondim}
\end{equation}
so that the non-dimensional complex potential is (from now on the
primes are omitted)
\begin{equation}
w(z)=z+\frac{1}{z}+2i\lambda\ln\frac{z(z-\zz)}{1-z\zzc}\,.
\label{nondim_potential}
\end{equation}
The $x$- and $y$-components of the fluid velocity are,
respectively, $u={\rm Re}\,\{dw/dz\}$ and $v=-{\rm
Im}\,\{dw/dz\}$.

The flow potential~(\ref{nondim_potential}) yields the following
Lagrangian equations of motion of two vortex points, $\zz(t)$ and
$\zzz(t)$:
\begin{equation}
\frac{dx_j}{dt}={\rm Re}\,\{f_j(\zz,\,\zzz)\}, \quad
\frac{dy_j}{dt}={\rm Im}\,\{f_j(\zz,\,\zzz)\}, \label{eqs_motion}
\end{equation}
where $j=1,\,2$, and
\begin{equation}
f_1(\zz,\,\zzz)=1-\frac{1}{\zz^2}-2i\lambda\biggl\{\frac{1}{\zz-\zzz}
+\frac{1}{\zz(1-\zz\zzzc)}\biggr\}\,,
\label{f1}
\end{equation}
\begin{equation}
f_2(\zz,\,\zzz)=1-\frac{1}{\zzz^2}+2i\lambda\biggl\{\frac{1}{\zzz-\zz}
+\frac{1}{\zzz(1-\zzz\zzc)}\biggr\}\,.
\label{f2}
\end{equation}

\section{Stationary positions of point vortices} \label{Stationary}

In relation with the experimental observations described in
Ref.~\cite{VanSciverNat}, we will now determine whether there
exist stationary locations of the vortex-antivortex pair, both
downstream and upstream in the reference frame of the disk. On the
complex plane, such stationary positions can only be the complex
conjugate points, say, $\zzo$ in the upper, and $\zzzo=\zzoc$ in
the lower half-plane, respectively. Since the imposed flow is in
the positive $x$-direction, the velocities of the both vortex
points can be zero only in the case where the vortices located at
$\zzo$ and $\zzzo$ have the negative and the positive polarity,
respectively. This implies $\Gamma<0$, so that the non-dimensional
parameter $\lambda$ defined by the last relation~(\ref{nondim}) is
positive. To ensure that the point vortices located at $\zzo$ and
$\zzzo$ are stationary, it is sufficient to require that the
complex velocity at $\zzzo=\zzoc$ is zero, i.e.
$f(\zzo)=f_2(\zzo,\,\zzoc)=0$, where the function $f_2$ is defined
by relation~(\ref{f2}). Introducing the function
$F(x,\,y)=yz^*(1-(z^*)^2)f(z)$, we obtain the following pair of
coupled equations for the coordinates of the stationary point:
\begin{equation}
{\rm Im}\,F(x_1^0,\,y_1^0)=0, \quad {\rm Re}\,F(x_1^0,\,y_1^0)=0,
\label{ImRe}
\end{equation}
where
\begin{equation}
{\rm Im}\,F(x,\,y)=4xy(y-\lambda)(x^2-y^2-1), \label{ImF}
\end{equation}
\begin{equation}
{\rm
Re}\,F(x,\,y)=(x^2-y^2)(1-x^2+y^2)(y-\lambda)+4x^2y^2(y-\lambda)-y(1-x^2+y^2)+2\lambda
y^2. \label{ReF}
\end{equation}
Consider first Eq.~(\ref{ImF}). Obviously, $y_1^0\neq0$. For the
root $y_1^0=\lambda$, from Eq.~(\ref{ReF}) we find
$x_1^0=\pm\sqrt{1+\lambda^2}$, so that $(x_1^0)^2+(y_1^0)^2=1$;
this solution corresponds to the vortex points on the disk
boundary and should be discarded.

\begin{figure}
\begin{tabular}[b]{cc}
\includegraphics[height=0.33\linewidth]{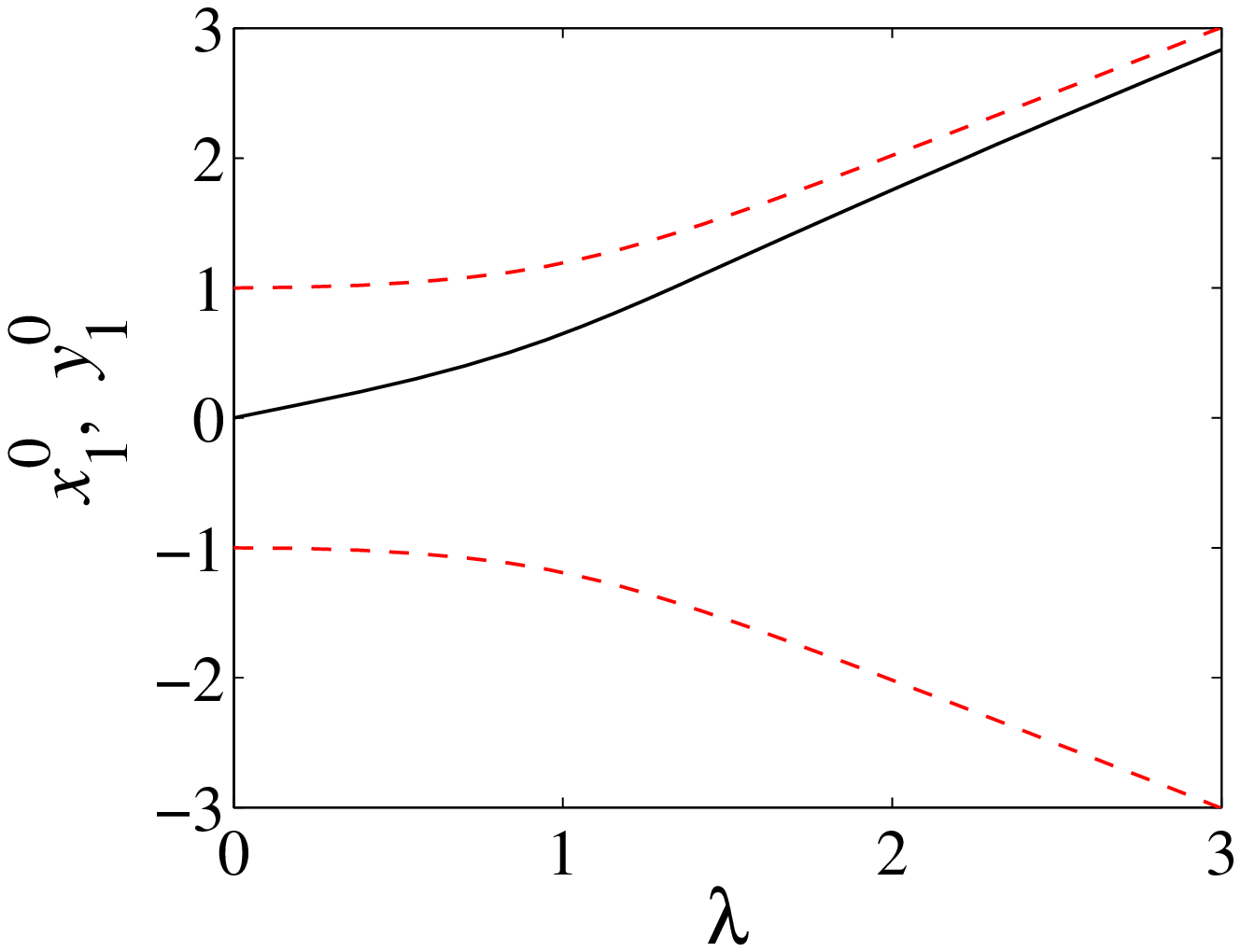}&
\includegraphics[height=0.33\linewidth]{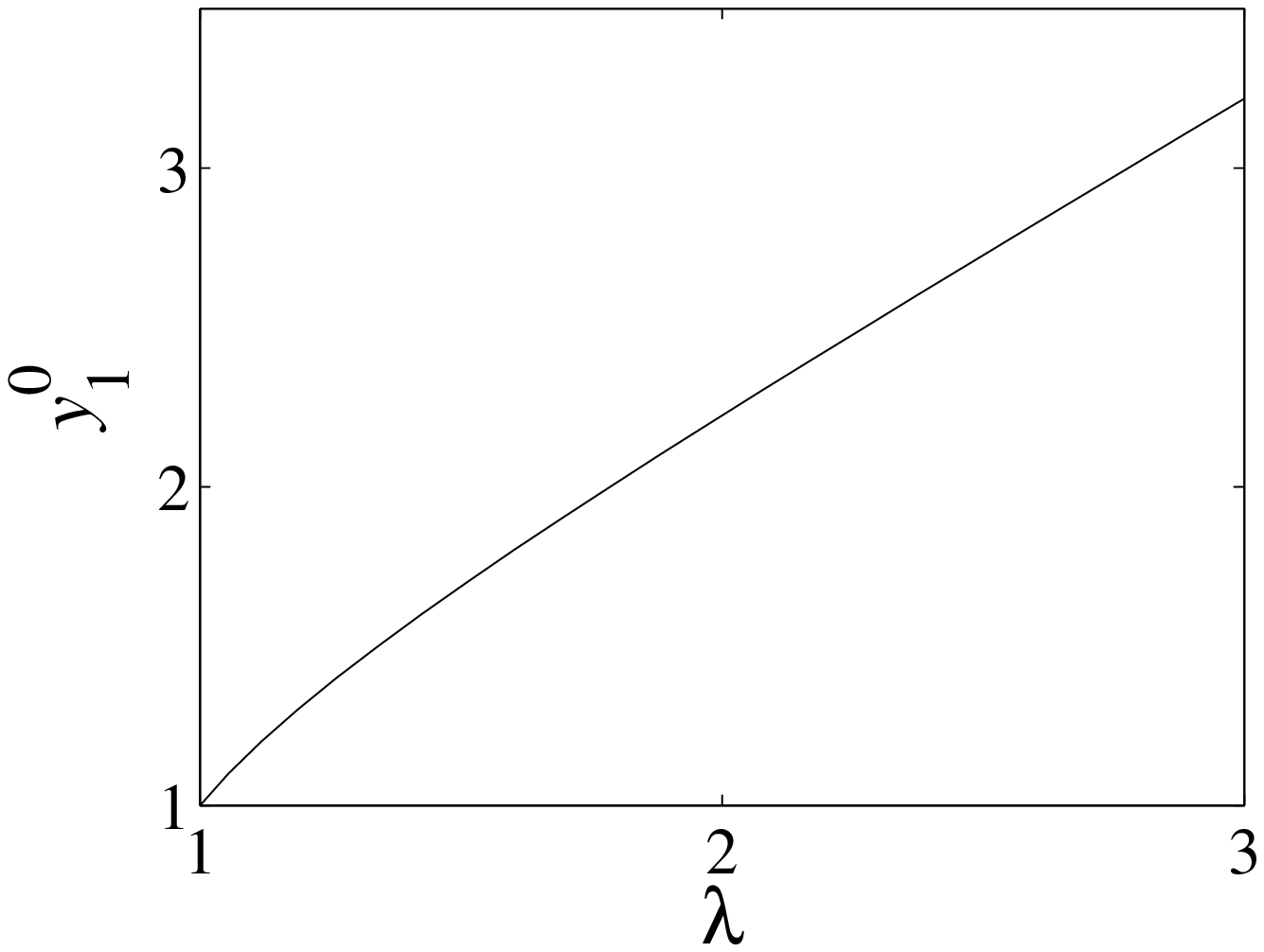}\\
\end{tabular}
\caption{(Color online). Left: coordinates $x_1^0$ -- dashed (red)
lines, and $y_1^0$ -- solid (black) line of the family~{\it
1}$^{\rm o}$ stationary points in the upper half-plane as
functions of the non-dimensional circulation $\lambda$. Right:
coordinates of the family {\it 2}$^{\rm o}$ stationary vortex
point in the upper half-plane ($x_1^0=0$).} \label{fig:2}
\end{figure}

This leaves two families of roots corresponding to {\it 1}$^{\rm
o}$: $1-(x_1^0)^2+(y_1^0)^2=0$, and {\it 2}$^{\rm o}$:~$x_1^0=0$.
For family {\it 1}$^{\rm o}$ Eq.~(\ref{ReF}) reduces to
$(y_1^0)^3-\lambda(y_1^0)^2+(y_1^0)-\lambda/2=0$. Solution of this
equation yields the coordinates $x_1^0$ and $y_1^0$, shown in
Fig.~\ref{fig:2} (left), of stationary locations of the vortex
points. Note the asymptotic behavior, $x_1^0\sim\pm\lambda$ and
$y_1^0\sim\lambda$ as $\lambda\to\infty$.

It can be seen that there exist stationary configurations of the
vortex-antivortex pair, both downstream and upstream from the
disk; such configurations are symmetric with respect to the
$y$-axis.

Family {\it 2}$^{\rm o}$ corresponds to the location of the
vortex-antivortex pair on the $y$-axis. Setting $x=0$,
Eq.~(\ref{ImF}) yields
$(y_1^0)^4-\lambda(y_1^0)^3+2(y_1^0)^2-3\lambda y_1^0+1=0$. For
$y_1^0\geq1$ (outside the disk) the real solution of this equation
is shown in Fig.~\ref{fig:2} (right). Note the asymptotic behavior
$y_1^0\sim\lambda$ as $\lambda\to\infty$.

\section{Apparent stability of stationary points} \label{stability}

The motion of two point vortices whose initial positions are
perturbed around the corresponding stationary values, i.e.
\begin{equation}
x_j(0)=x_j^0(\lambda)+(\Delta x_j)_0, \quad
y_j(0)=y_j^0(\lambda)+(\Delta y_j)_0, \label{perturbations}
\end{equation}
where $j=1,\,2$, $x_2^0(\lambda)=x_1^0(\lambda)$, and
$y_2^0(\lambda)=-y_1^0(\lambda)$, is governed by the system of
four Lagrangian equations~(\ref{eqs_motion}) for $x_1(t)$,
$y_1(t)$, $x_2(t)$, and $y_2(t)$. Numerical solution of
Eqs.~(\ref{eqs_motion}) shows that any perturbation leads in
general to sweeping of vortex points by the imposed, uniform at
infinity, flow around the disk, so that $x_j(t)\to+\infty$ as
$t\to\infty$ (the only exception being the case of symmetric
initial perturbations
such that $(\Delta x_1)_0=(\Delta x_2)_0$ and
$(\Delta y_1)_0=(\Delta y_2)_0$, in which case both vortex points
move along closed trajectories around stationary points).

\begin{figure}
\centering \epsfig{figure=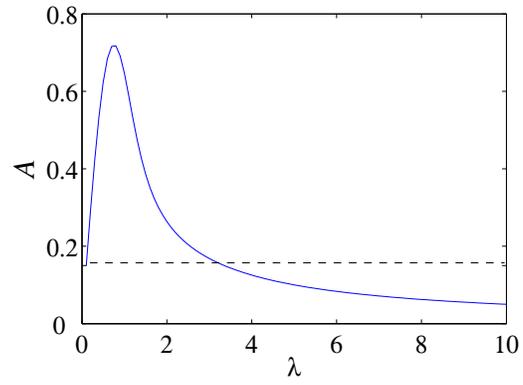,height=2.1in,angle=0}
\caption{(Color online). Function $A(\lambda)$. The dashed
horizontal line corresponds to $A=0.15$.} \label{fig:3}
\end{figure}

Of particular interest would be an analysis of motion of point
vortices during the first $t_1=70$ non-dimensional units of time
corresponding to the dimensional duration, $t_{\rm exp}=10\,{\rm s}$
of the experiment~\cite{VanSciverNat}
(see below Sec.~\ref{interpretation}). We will be
concerned with the values of $\lambda$ for which the vortex points,
starting their motion near the stationary points
$(x_1^0(\lambda),\,\pm y_1^0(\lambda))$, remain in the sufficiently
close vicinity of their initial positions for at least $t_1$
time units. A simple {\it a priory} estimate for
such values of $\lambda$ can be obtained from
Eqs.~(\ref{eqs_motion})-(\ref{f2}) as follows. At the stationary
positions, $\zz=\zzo(\lambda)$ and $\zzz=\zzoc$ we obviously have
$f_{1,\,2}(\zzo,\,\zzoc)=0$. If the vortex point $\zz(t)$ is located
initially in the close vicinity of $\zzo$, so that
$\vert\zz(0)-\zzo\vert\ll\vert\zzo\vert$, the magnitude of its
velocity at $t=0$ can be estimated, expanding
$f_1$ around the stationary point,
as $A(\lambda)\vert\zz-\zzo\vert$, where
\begin{equation}
A(\lambda)=\biggl\vert\frac{\partial
f_1(\zzo,\,\zzoc)}{\partial\zzo}\biggr\vert\,. \label{A}
\end{equation}
The function $A(\lambda)$ is shown in Fig.~\ref{fig:3}.

If, at $t=0$, the distance between the vortex $\zz$ and the
stationary point $\zzo$ is small, the distance traveled by the
vortex $\zz$ during time $t_1$ can be estimated as $l\sim
A(\lambda)\vert\zz(0)-\zzo\vert t_1$. Assuming the distance $l$ to
be not larger than few (say, 10) times the initial distance
between the vortex and the stationary point, we find that, during
the first $t_1=70$ non-dimensional units of time, the vortex point
will remain relatively close to its initial position provided
$A(\lambda)\lesssim0.15$; this corresponds to $\lambda\gtrsim3$,
see Fig.~\ref{fig:3}. Consequently, on a time scale corresponding
to $t_1$ non-dimensional units the vortices located sufficiently
close to the stationary points $\zzo(\lambda)$ and
$(\zzo(\lambda))^*$ will appear, for $\lambda\gtrsim3$, as
apparently stable.

\begin{figure}
\begin{tabular}[b]{cc}
\includegraphics[height=0.33\linewidth]{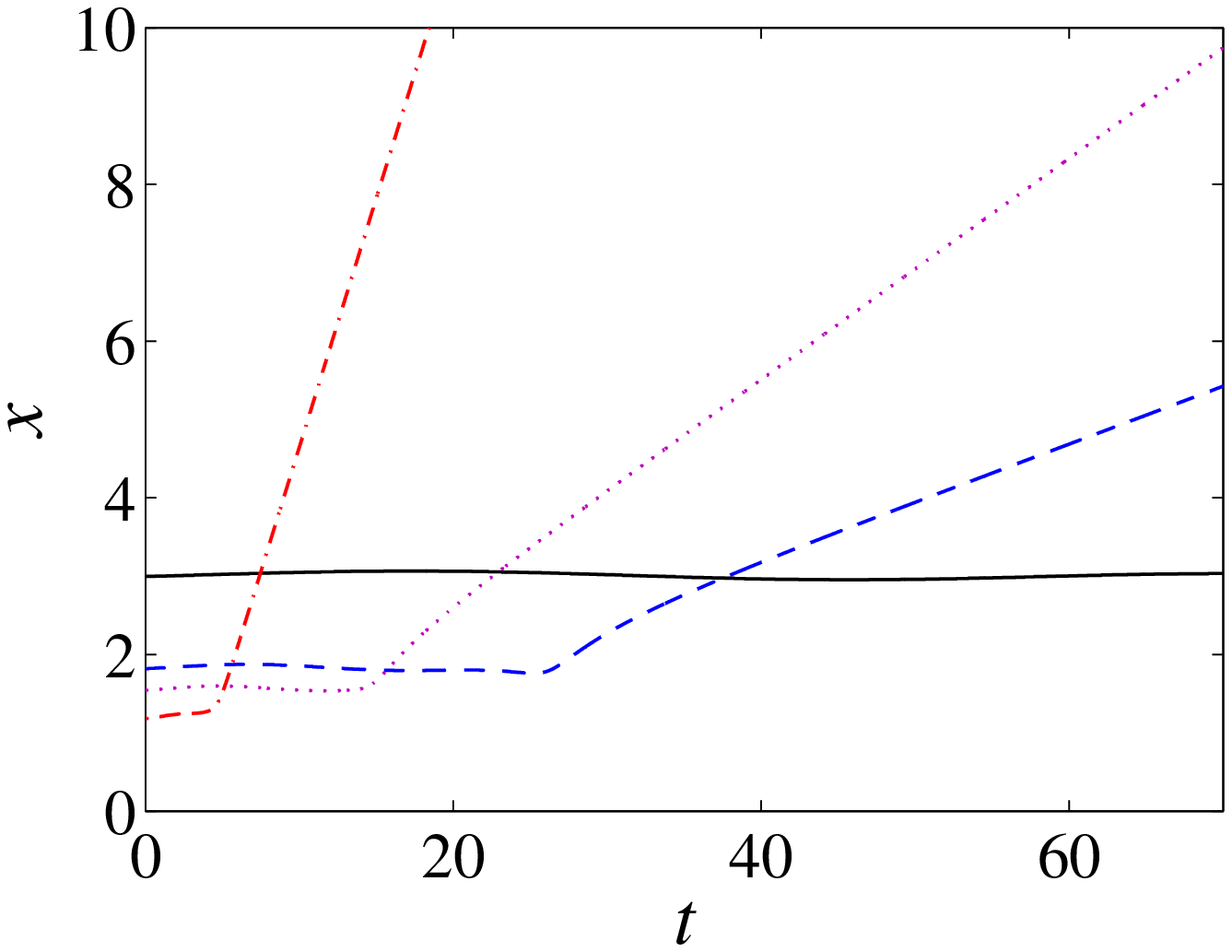}&
\includegraphics[height=0.33\linewidth]{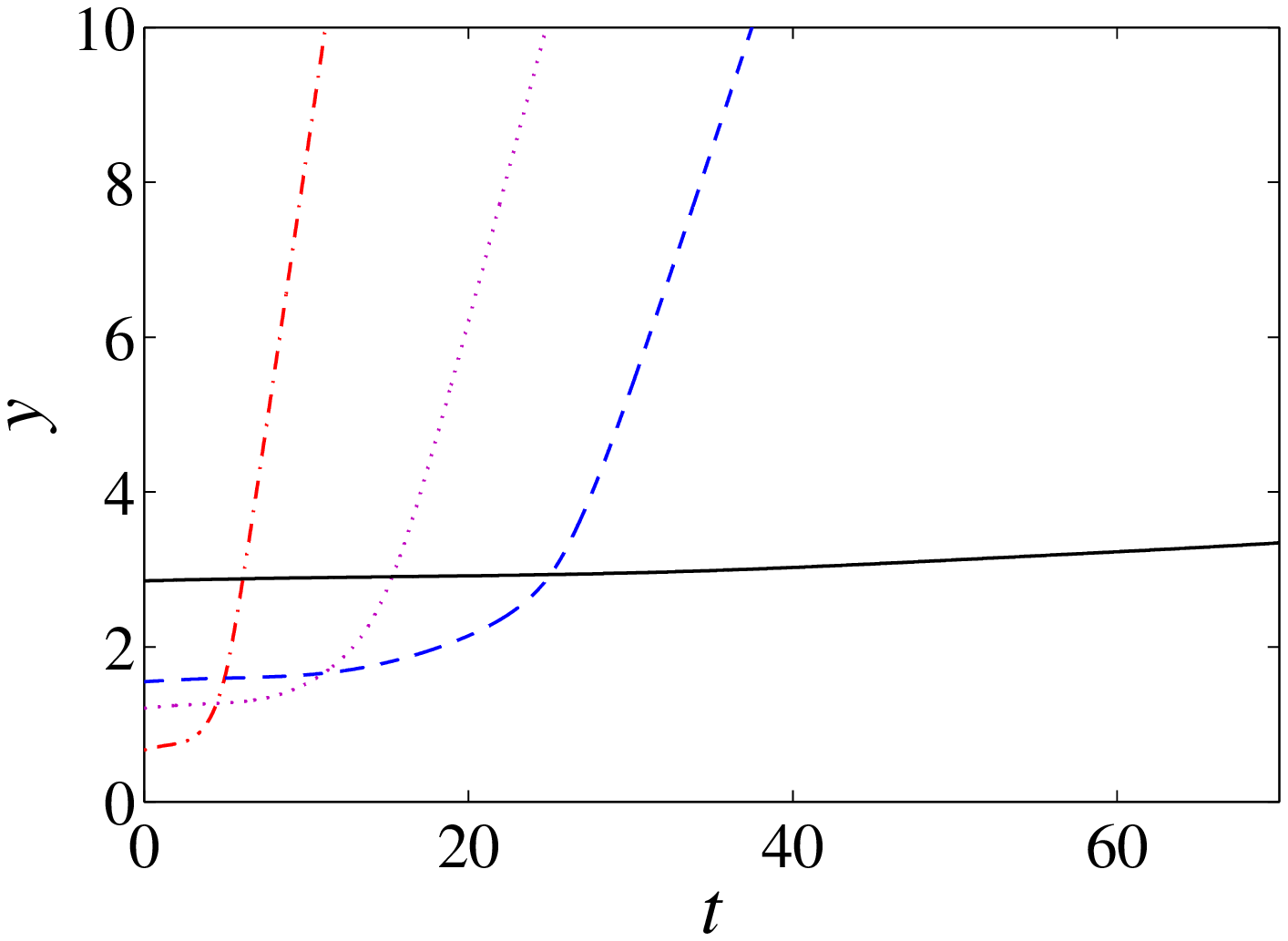}\\
\end{tabular}
\caption{(Color online). Time-dependent coordinates $x_1(t)$
(left) and $y_1(t)$ (right) of the vortex point initially located
in the vicinity of the stationary point
$(\xxo(\lambda),\,\yyo(\lambda))$ in the upper half-plane
downstream of the disk. Dot-dashed (red) lines: $\lambda=1$,
$\xxo\approx1.191$, $\yyo\approx0.648$; dotted (purple):
$\lambda=1.5$, $\xxo\approx1.554$, $\yyo\approx1.189$; dashed
(blue): $\lambda=1.8$, $\xxo\approx1.828$, $\yyo\approx1.531$;
solid (black): $\lambda=3$, $\xxo\approx3.005$,
$\yyo\approx2.834$. Initial positions of the vortex-antivortex
pair are defined by perturbations $(\Delta x_1)_0=-0.01$, $(\Delta
y_1)_0=0.02$, $(\Delta x_2)_0=-0.03$, and $(\Delta y_2)_0=-0.01$.}
\label{fig:4}
\end{figure}

\begin{figure}
\begin{tabular}[b]{cc}
\includegraphics[height=0.33\linewidth]{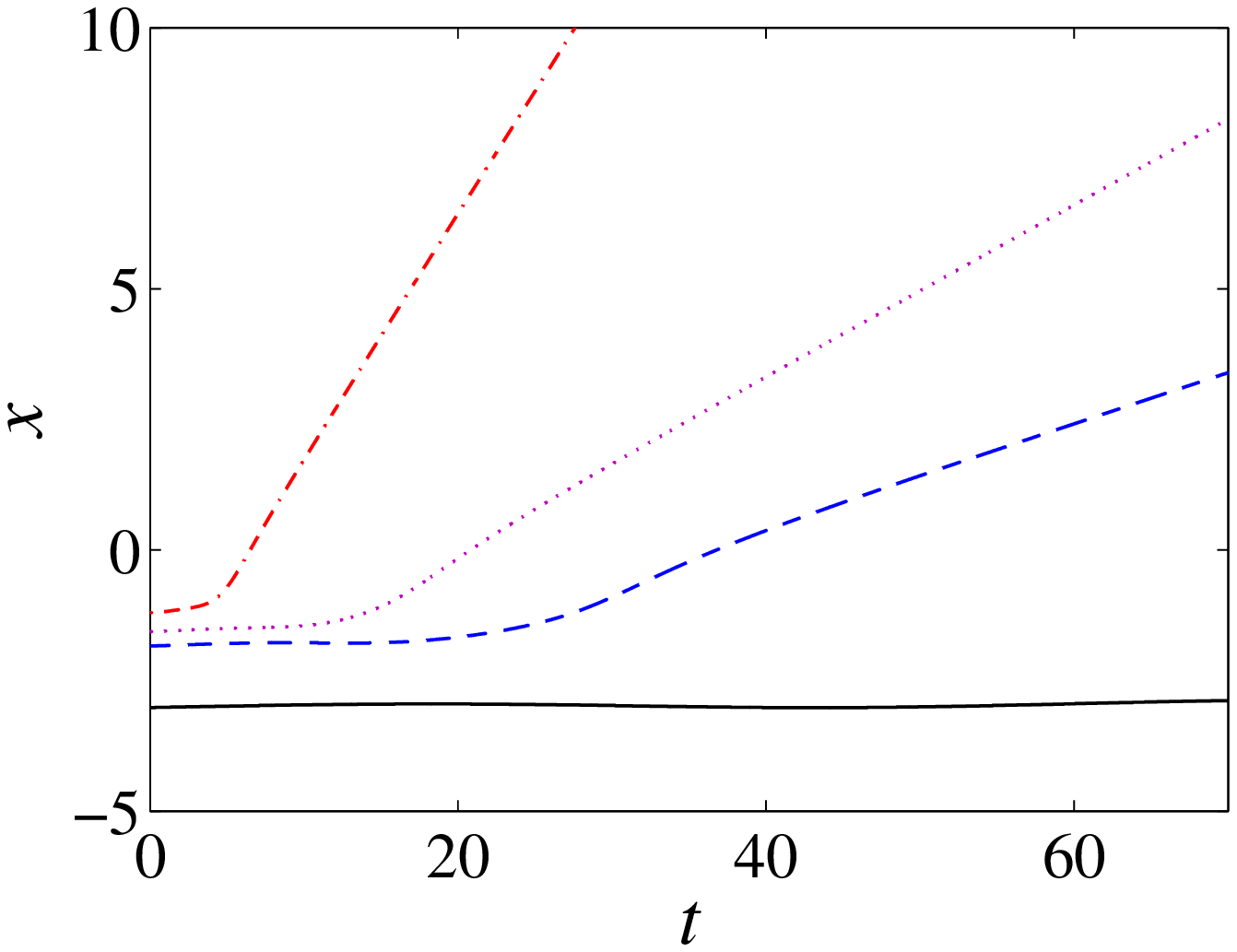}&
\includegraphics[height=0.33\linewidth]{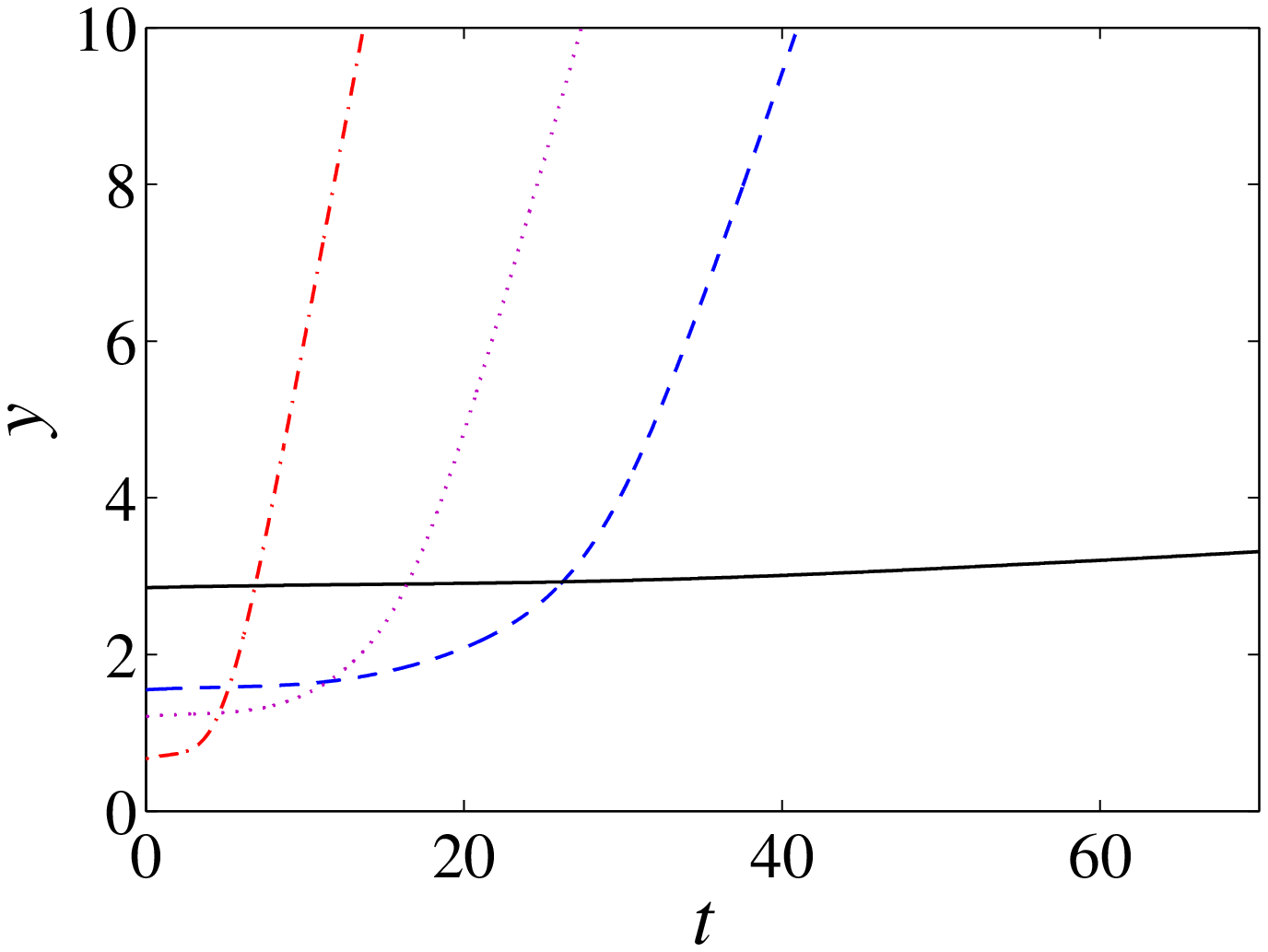}\\
\end{tabular}
\caption{(Color online). Coordinates of the vortex point initially
located in front of the disk. Values of $\lambda$,
$\yyo(\lambda)$, $\vert\xxo(\lambda)\vert$, $(\Delta x_j)_0$, and
$(\Delta y_j)_0$ are the same as for Fig.~\ref{fig:5}.}
\label{fig:5}
\end{figure}

\begin{figure}
\begin{tabular}[b]{cc}
\includegraphics[height=0.33\linewidth]{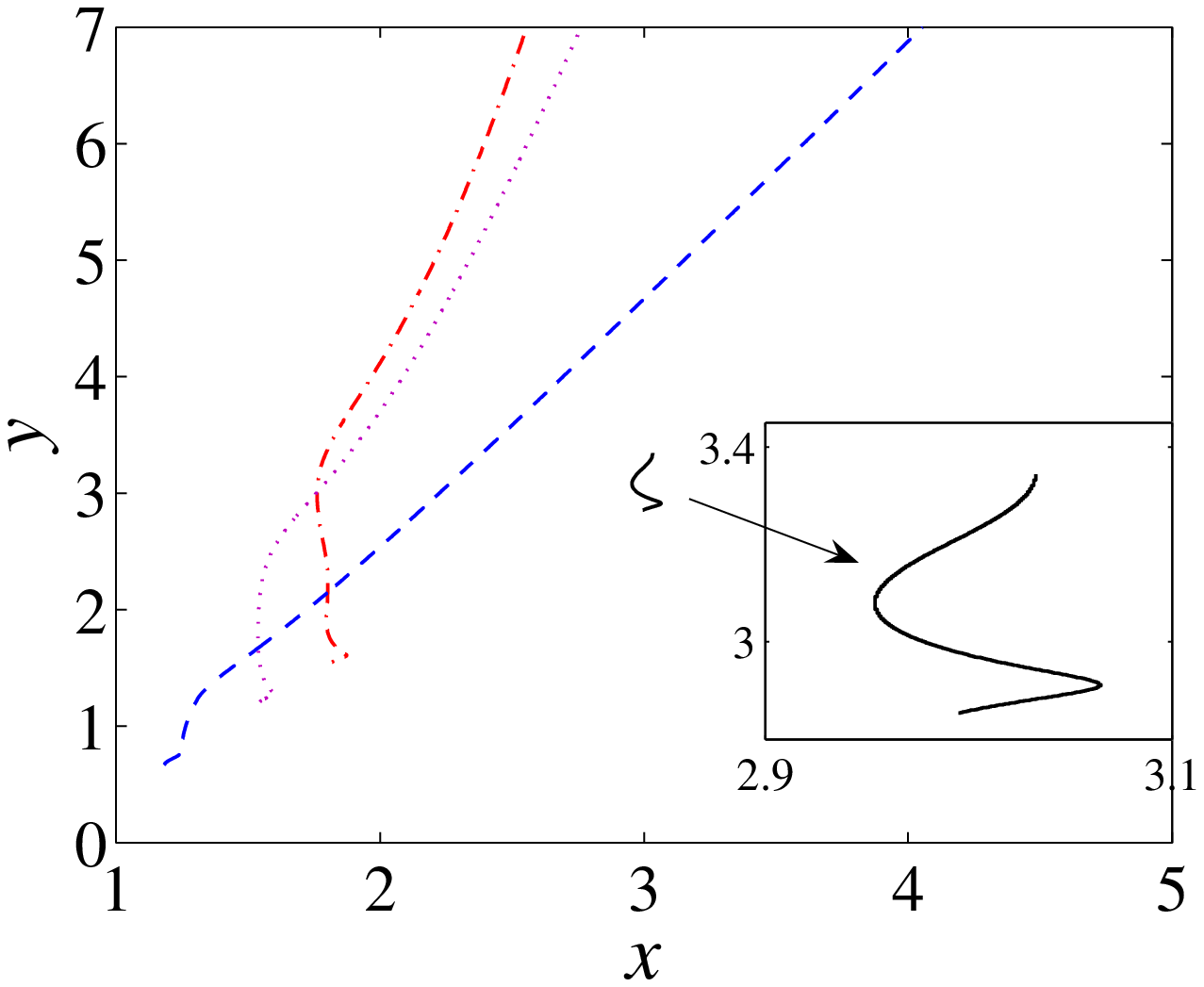}&
\includegraphics[height=0.33\linewidth]{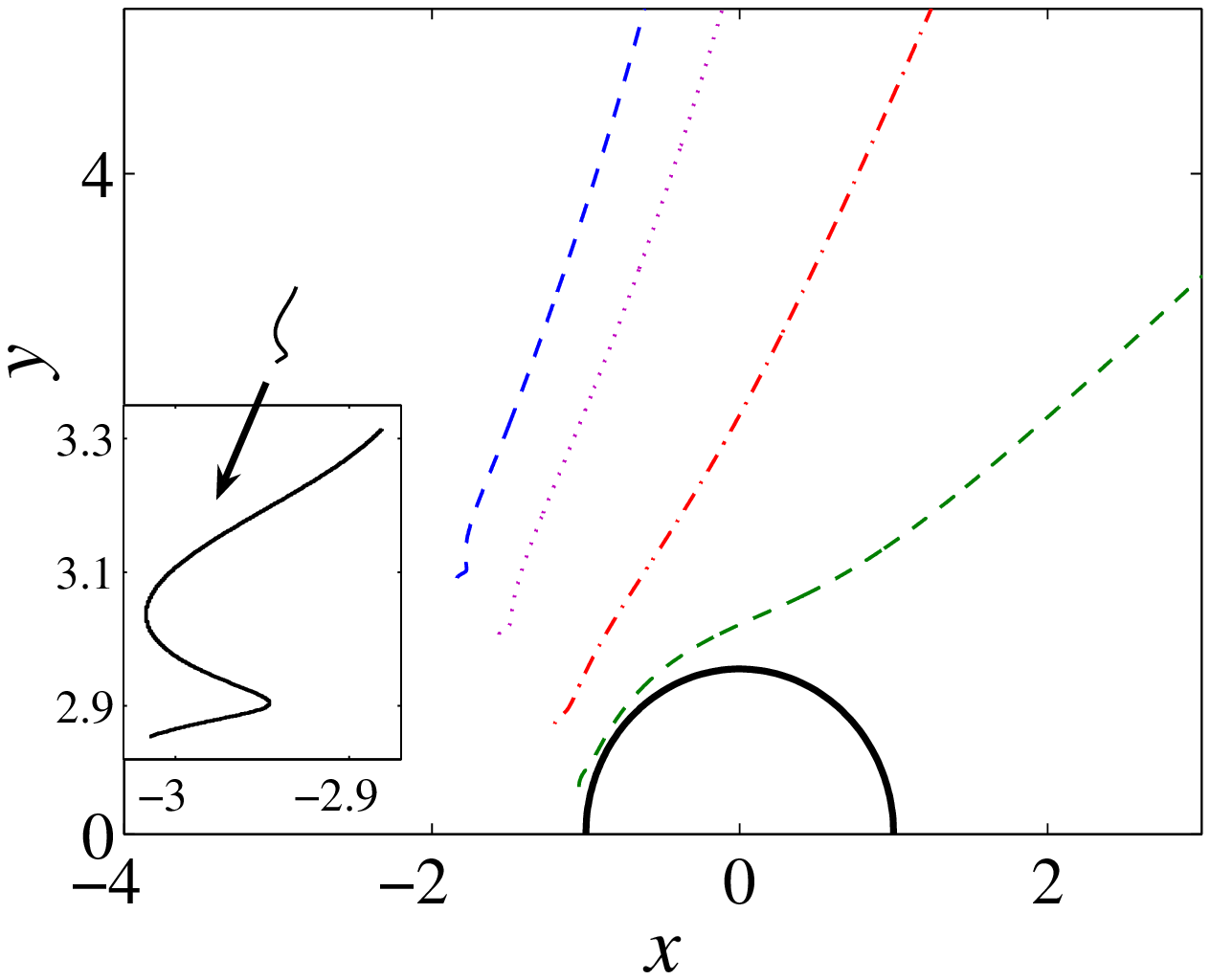}\\
\end{tabular}
\caption{(Color online). Trajectories of the vortex point. Left:
trajectories corresponding to Fig.~\ref{fig:4}. Right: first four
trajectories, from left to right, correspond to Fig.~\ref{fig:5};
the last trajectory corresponds to $\lambda=0.5$
($\xxo(0.5)\approx-1.035$, $\yyo(0.5)\approx0.266$). Inserts show
the trajectories for $\lambda=3$ during $t_1=70$ non-dimensional
time units.} \label{fig:6}
\end{figure}

This conclusion is illustrated by the following example. For
initial positions defined by the perturbations $(\Delta
x_1)_0=-0.01$, $(\Delta y_1)_0=0.02$, $(\Delta x_2)_0=-0.03$, and
$(\Delta y_2)_0=-0.01$, the results of numerical calculation of
the motion of vortex points are shown in Figs.~\ref{fig:4} and
\ref{fig:6} (left) for vortices initially located at the rear of
the disk, and Figs.~\ref{fig:5} and \ref{fig:6} (right) for
vortices initially positioned in front of the disk. Our
calculations show that, for vortices initially located both at the
rear and in front of the disk, the period of time during which the
vortex points remain close to their initial locations increases
with the non-dimensional circulation $\lambda$. For $\lambda=3$
the magnitudes of displacement of vortices from their initial
locations, $\Delta r_j=[(x_j-x_j^0)^2+(y_j-y_j^0)^2]^{1/2}$, where
$j=1,\,2$, remain smaller than 0.44 (or less than 15\% of the
distance from the center of cylinder) for $t_1=70$ time units.
This result will be used in Sec.~\ref{interpretation} for
explanation of the apparent stability of eddies observed in the
experiment~\cite{VanSciverNat}. (During the same time interval, a
more pronounced displacement such that $\Delta r_j\sim1$ occurs
already for $\lambda=2.8$.) Note that this conclusion is not a
consequence of the specific choice of initial conditions but
remains quantitatively valid for any sufficiently small (such that
$\vert(\Delta x_j)_0\vert<0.03$ and $\vert(\Delta
y_j)_0\vert<0.03$) initial perturbation.

Similar results were also obtained for the motion of point
vortices in the vicinity of family {\it 2}$^{\rm o}$ stationary
points located at the $y$-axis. We found that the point vortices,
positioned initially near these stationary points, will remain in
their close vicinity during the non-dimensional time $t_1=70$
for considerably larger, compared with those
for
family {\it 1}$^{\rm o}$, values of the
non-dimensional circulation, i.e. for $\lambda\gtrsim7$; such values
of $\lambda$ correspond to the stationary points located at
$\yyo(\lambda)\gtrsim7$. Since in the
experiment~\cite{VanSciverNat} the boundary of the flow domain in
the $y$-direction was at the distance 6.25 cylinder radii,
no apparently stationary flow structures could be observed
corresponding to family {\it 2}$^{\rm o}$ stationary points.
Therefore, for the purpose of this work, the further, more
detailed,
analysis of the motion of point vortices in the vicinity of
family {\it 2}$^{\rm o}$ stationary points would be irrelevant.

The numerical results described in this Section can explain,
entirely from the classical fluid dynamics viewpoint without
invoking the mechanism of interaction between the normal
fluid and superfluid vortices, the apparent
stability of the normal eddies observed by Zhang and Van
Sciver \cite{VanSciverNat} both at the rear and in front of
the cylinder in the thermal counterflow, see
Sec.~\ref{interpretation} below.

\section{Interpretation of experimental observations.
Discussion and conclusions} \label{interpretation}

In the experiment~\cite{VanSciverNat} the velocity field was
recorded of a large number of micron-size particles injected in
turbulent He~II thermal counterflow. In classical viscous fluids,
provided a particle is sufficiently small, the viscous drag force
exerted on the particle dominates all other forces (such as the
inertial and added mass force, Saffman and Magnus lift forces, the
Basset memory force, etc.), so that the velocity field of a large,
dilute ensemble of solid particles faithfully maps the fluid flow
field. In turbulent He~II, due to interactions between particles and
quantized vortices, particles generally map neither the normal fluid
nor superfluid, as was demonstrated by recent
experimental~\cite{VanSciver,Bewley,Paoletti} and
theoretical~\cite{Poole,Sergeev,Kivotides} studies. However, in
Zhang and Van Sciver counterflow experiment~\cite{VanSciverNat} a
tangle of superfluid vortices was rather dilute, with the mean
intervortex distance $\ell\approx6\,\mu{\rm m}$ exceeding
substantially the particle diameter $d_{\rm p}=1.7\,\mu{\rm m}$.
This enables us to assume that close encounters between particles
and quantized vortices (and, in particular, the events of particle
trapping on quantized vortex cores) were relatively rare and,
therefore, the particulate flow field recorded in the
experiment~\cite{VanSciverNat} maps, at least qualitatively, the
velocity field of the normal fluid.

Of particular interest are the apparently stationary normal fluid
eddies observed by Zhang and Van Sciver both behind and in front
of the cylinder. Unlike the familiar
Von K\'arm\'an vortex street shed by a cylinder, such structures
were never observed in the classical
viscous flow. In Ref.~\cite{VanSciverNat} Zhang and Van Sciver
attributed the existence of apparently stationary normal eddies
to the mutual friction interaction between quantized vortices and
the normal fluid.

In this work we argue that, based on the idealized flow model
considered
above in Secs.~\ref{motion}-\ref{stability}, the experimental
results~\cite{VanSciverNat} can be interpreted without invoking
the mechanism of interaction between the normal fluid and
quantized vortices. It must be emphasized that the flow analyzed
in Secs.~\ref{motion}-\ref{stability} is that of the inviscid
fluid, while the normal flow in the experiment~\cite{VanSciverNat}
is obviously viscous, so that the following arguments and
estimates should be regarded as qualitative. However, in the
experiment~\cite{VanSciverNat} the Reynolds numbers defined by
the diameter of the cylinder were at least of the order of
$2\times10^4$, so that the considered inviscid, potential flow
can be used as a reasonable approximation to
a distribution of the Reynolds averaged velocity of the turbulent
normal fluid around the cylinder.

In the experiment described in Ref.~\cite{VanSciverNat} the
large-scale normal eddies, both downstream and upstream of the
cylinder, were observed at a distance about 3 cylinder radii at
the angles $\pm45^{\rm o}$ and $\pm135^{\rm o}$ to the axis
through the center of cylinder in the direction
of the undisturbed normal flow. For the normal fluid velocity
$\vn\approx2.2\,{\rm cm/s}$, these circulation patterns appear as
stable for the duration of the experiment
$t_{\rm exp}\approx10\,{\rm s}$.

Our calculation of motion of point vortices in the imposed
potential flow around the disk showed that there exist stationary
locations of point vortices, both downstream and upstream of the
disk. These locations are unstable: any perturbation of the
initial stationary
positions of point vortices leads, eventually, to sweeping of
point
vortices away from their initial locations. The distance between
the stationary points and the center of the disk increases with
the non-dimensional circulation $\lambda=-\Gamma/(4\pi Ua)$. Also
increases with $\lambda$ the period of time during which the
slightly perturbed vortex points remain in the close vicinity of
the corresponding stationary points. For $\lambda\gtrsim3$, in
the case where the magnitudes of initial perturbations of
coordinates of stationary positions are smaller than 0.03,
we found that the vortices will remain in the
close vicinity of the stationary points for at least $t_1=70$
non-dimensional units of time. In the case where the undisturbed
fluid velocity, $U$ and the disk radius, $a$ are identified,
respectively, with
the normal fluid
velocity, $\vn\approx22\,{\rm mm/s}$ and the cylinder radius,
$3.175\,{\rm mm}$ in the
experiment~\cite{VanSciverNat}, the corresponding dimensional
time is $t=t_1a/U\approx10\,{\rm s}\approx t_{\rm exp}$.
Therefore, on a time scale corresponding to the duration of
experiment~\cite{VanSciverNat}
the vortices will remain close to
the positions of the stationary points and, hence,
appear as stable. Moreover, for
$\lambda=3$ the
non-dimensional coordinates of the stationary points,
$x_j^0\approx\pm3.005$ and $y_j^0\approx\pm2.834$ correspond to the
locations, in the cited experiment, of the observed
apparently stable normal
eddies at the angles $\pm45^{\rm o}$ and $\pm135^{\rm o}$ to the
axis through the center of the cylinder in the direction of the
undisturbed normal flow.

In summary, having considered the motion of the vortex-antivortex
pair in the two-dimensional Euler flow around the disk, we found
stationary solutions which enabled us to interpret the existence
of apparently stale normal eddies in the
experiment~\cite{VanSciverNat} without invoking interactions
between the normal fluid and quantized vortices. Nevertheless, the
mutual friction between quantized vortices and the normal fluid,
together with the polarization of the vortex tangle, might be
responsible for the emergence of normal eddies in the first place:
the polarized cluster of superfluid vortices would rotate the
normal fluid (see e.g. Refs.~\cite{Vinen,Barenghi}) which, in
turn, would drag along the tracer particles in the PIV experiment.
(On a related problem, it should be mentioned that the results of
H\"anninen {\it et al.}~\cite{Hanninen} are hinting at the
possibility of formation of a classical wake of quantized vortices
behind an oscillating sphere. It is not yet known whether a
similar phenomenon occurs in the flow past a cylinder, although
the experimental results~\cite{VanSciverNat} seem to point in this
direction.) Assuming that, at least within the domains occupied by
the circulation cells observed in the
experiment~\cite{VanSciverNat}, the normal fluid and the
superfluid are fully interlocked through the mutual friction, it
is easy to estimate that, for the parameters typical of the cited
experiment, the value of the non-dimensional circulation,
$\lambda=3$ corresponds to $N\approx2.6\times10^4$ quanta of
circulation, $\kappa=10^{-3}\,{\rm cm^2/s}$. Considering a cluster
of $N$ polarized vortices with a cross-section of radius $a_{\rm
cl}$, we find that its (polarized) vortex line density would be
$L'\approx8.3\times10^4\,{\rm cm^{-2}}$, which is a small, 8\%
polarization of the total (random) vortex line density,
$L_0\approx10^6\,{\rm cm^{-2}}$ typical of the
experiment~\cite{VanSciverNat}. Therefore, as envisaged in
Refs.~\cite{Vinen,Barenghi}, even a small polarization of the
vortex tangle would be sufficient to generate normal circulation
patterns which, if located near the stationary points of the
considered Euler flow, can exist as apparently stable for the
duration of experiment.

\section{Acknowledgments}

We are grateful to S.~W.~Van Sciver, W.~F.~Vinen, and L.~Skrbek
for fruitful discussions.

\vfill
\eject

\newpage

%%%%%%%%%%%%%%%%%%%%%%%%%%%%%%%%%%%%%%%%%%%%%%%%%%%%%%%%%%%%%%%
% REFERENCES

\newpage

%%%%%%%%%%%%%%%%%%%%%%%%%%%%%%%%%%%%%%%%%%%%%%%%%%%%%%%%%%%%%%%
% FIGURES
%%%%%%%%%%%%%%%%%%%%%%%%%%%%%%%%%%%%%%%%%%%%%%%%%%%%%%%%%%%%%%%

% FIGURE 1

\clearpage

% FIGURE 2

\clearpage

% FIGURE 3

\clearpage

% FIGURE 4

\clearpage

% FIGURE 5

\clearpage

% FIGURE 6

\end{document}